\documentclass[showpacs, pra,onecolumn,preprintnumbers ,amsmath, amssymb, superscriptaddress, aps]{revtex4-2}
\usepackage{color}
\usepackage{amsmath,amssymb}
\usepackage{pifont}
\usepackage{amssymb}  
\usepackage{bbold}
\usepackage{float}
\usepackage{subfloat}
\usepackage[squaren]{SIunits}

\usepackage[caption=false]{subfig}
\usepackage{tikz}
\usepackage{makecell}
\usepackage{subfig}
\usepackage{pifont}   
\usepackage{graphicx} 
\graphicspath{{Figuress/}}
\usepackage{dcolumn}  
\usepackage{bm}       
\usepackage{multirow} 
\usepackage{placeins}
\usepackage[colorlinks]{hyperref}
\usepackage{mathtools}
\usepackage{appendix}

\begin{document}
	
	\title{  
		Quantum steering and entanglement for coupled systems: exact results}
	\date{\today}

	\author{Radouan Hab arrih}
	\email{habarrih46@gmail.com}
	\affiliation{Label excellence, FAR avenue, Main road Fez, Meknes 50000, Morocco}
	\affiliation{Laboratory of Theoretical Physics, Faculty of Sciences, Choua\"ib Doukkali University, PO Box 20, 24000 El Jadida, Morocco}
	\author{Ayoub Ghaba}
	\affiliation{Laboratory of Nanostructures and Advanced Materials, Mechanics and Thermofluids, Faculty of Sciences and Techniques, Hassan II
		University, Mohammedia, Morocco}
	\author{Ahmed Jellal}
	\email{a.jellal@ucd.ac.ma}
	\affiliation{Laboratory of Theoretical Physics, Faculty of Sciences, Choua\"ib Doukkali University, PO Box 20, 24000 El Jadida, Morocco}

	\pacs{03.65.Fd, 03.65.Ge, 03.65.Ud, 03.67.Hk\\
		{\sc Keywords:} Phase space, excitation, fluctuations, steering, entanglement, asymmetry, resonance}

	\begin{abstract}
		
		Using the Wigner function in phase space, we study quantum steering and entanglement between two coupled harmonic oscillators. We derive expressions for purity and quantum steering in both directions and identify several important selection rules. Our results extend the work reported in {\color{blue} [Phys. Rev. E 97, 042203 (2018)]} focused on the weak coupling regime, revealing significant deviations in the ultra-strong coupling regime. In particular, Makarov's prediction of a separable ground state contrasts with our exact calculations, highlighting the limitations of his approach under strong coupling conditions. We show that quantum steering between excited oscillators is completely absent even in the ultra-strong coupling regime. Similarly, resonant oscillators have no steering, and ground states cannot steer any receiver state. We find that quantum steering becomes notably more pronounced as the system approaches resonance and within specific ranges of ultra-strong coupling. This behavior is marked by a clear asymmetry, where steering is present in only one direction, highlighting the delicate balance of interaction strengths that govern the emergence of quantum correlations. These results advance our understanding of how excitation levels and coupling strengths influence quantum steering and entanglement in coupled harmonic oscillators.

	\end{abstract}

	\maketitle
	
\section{Introduction}

Quantum steering is a fascinating quantum phenomenon that has attracted considerable attention in the fields of quantum information theory and quantum mechanics. The concept was originally introduced by Schrödinger \cite{intro1} as a response to the Einstein-Podolsky-Rosen (EPR) paradox \cite{intro2}, which challenged the completeness of quantum mechanics by suggesting the possibility of "spooky action at a distance." Quantum steering describes a situation where one party, typically referred to as the "sender," can non-locally influence the state of another party, the "receiver," through local measurements made on its own subsystem. This non-local influence cannot be explained by classical models based on local hidden variables, distinguishing it from classical correlations and even from entanglement.
The formalism of quantum steering was further extended by Wiseman, Jones, and Doherty \cite{intro3}, who positioned it between quantum entanglement and Bell nonlocality. Unlike entanglement, which requires strong correlations between distant systems, quantum steering can occur even when systems are not fully entangled. This property makes quantum steering a unique type of quantum correlation that shows nonlocal effects but is not as extreme as Bell nonlocality, which arises from violations of Bell's inequalities. Quantum steering is demonstrated when the measurement outcomes on a subsystem cannot be explained by any local hidden state model, indicating that the correlations are purely quantum and not due to classical interactions.

In recent studies, quantum steering has been recognized for its profound implications in various quantum technologies, including quantum information processing and quantum cryptography. Its unique ability to demonstrate non-locality in a less stringent form than Bell nonlocality makes it an essential tool for understanding quantum correlations in practical applications. Quantum steering has been shown to have potential applications in secure communication, where it could be used for developing new cryptographic protocols that take advantage of quantum mechanics’ inherent unpredictability and non-locality \cite{intro4, intro5, intro6}. Furthermore, it provides critical insights into the broader understanding of quantum non-locality, offering a more nuanced view of how quantum systems can exhibit correlations that transcend classical physics and challenge our perception of locality in the universe \cite{intro7, intro8, intro9}.


Coupled harmonic oscillators are fundamental systems in quantum mechanics that exhibit a variety of well-known quantum properties. Due to their relatively simple and well-defined mathematical framework, they have become a powerful and versatile tool for modeling and analyzing a wide range of physical systems \cite{intro10}. Their quantum states can exhibit rich and complex behaviors, including the generation of quantum entanglement, the development of various quantum correlations, and other non-classical phenomena that are crucial for the advancement of quantum technologies \cite{intro11, intro12, intro13, intro14, intro15, Makarov1, Makarov2}. These interactions lead to the emergence of different quantum states, which are generally classified into two types: Gaussian and non-Gaussian states. Gaussian states, which are fully described by second-order moments of the system's quadrature operators, are commonly found in quantum harmonic oscillator systems. Non-Gaussian states, on the other hand, exhibit more complicated correlations and are typically associated with more exotic quantum behaviors. While much research has focused on understanding the properties of the ground states in coupled oscillators, recent studies suggest that the excited states play an even more important role in generating quantum correlations. These excited states, due to their higher energy and more complex dynamics, often serve as a richer source of entanglement and other quantum effects, making them essential for exploring the full range of quantum phenomena in these systems.

In \cite{Makarov1}, a method was introduced to analytically determine the Schmidt modes for coupled quantum harmonic oscillators. However, this approach relied on certain approximations that overlooked the contributions of Gaussian quantum entanglement, leading to an inaccurate result where the von Neumann entropy vanishes ($S_v(\Psi_{(0,0)}) = 0$). This approximation, while simplifying the calculations, fails to account for the full complexity of the entanglement structure in these systems. In contrast, our study revisits the problem of quantum entanglement in coupled oscillators by employing the purity function in phase space, a technique that avoids the need for such approximations. By using this more precise method, we are able to capture the true nature of quantum correlations and entanglement without neglecting any of the critical components. Furthermore, we extend the analysis by investigating quantum steering between the oscillators, which provides additional insight into the underlying quantum dynamics. Through this, we identify specific selection rules for quantum numbers and physical parameters that are crucial for sustaining quantum steering. These findings contribute to a more comprehensive understanding of the conditions under which quantum steering can be preserved, offering valuable insights for both theoretical studies and practical applications in quantum information processing and communication.


The structure of the paper is as follows. In Sec. \ref{sec1} we introduce the coupled oscillator system, give details of its physical setup, and derive the corresponding energy spectrum. In Sec. \ref{sec3}, we determine the Wigner function used to compute the phase space fluctuations. These will be used to determine the Heisenberg uncertainties and virtual excitations in Sec. \ref{sec4}. In Sec. \ref{sec5} we explore the quantum entanglement properties of the system, using the phase space formalism to study how entanglement arises and evolves in this context. In Sec. \ref{sec6}, we consider quantum steering and focus on its stationary properties within the system. We analyze the conditions that give rise to quantum steering and identify factors that affect its persistence. Finally, in Sec. \ref{sec7}, we summarize our main findings and suggest possible avenues for future research.

\section{Energy spectrum \label{sec1}}

\begin{figure}[H]\label{f1}
\centering
\includegraphics[width=9cm, height=7cm]{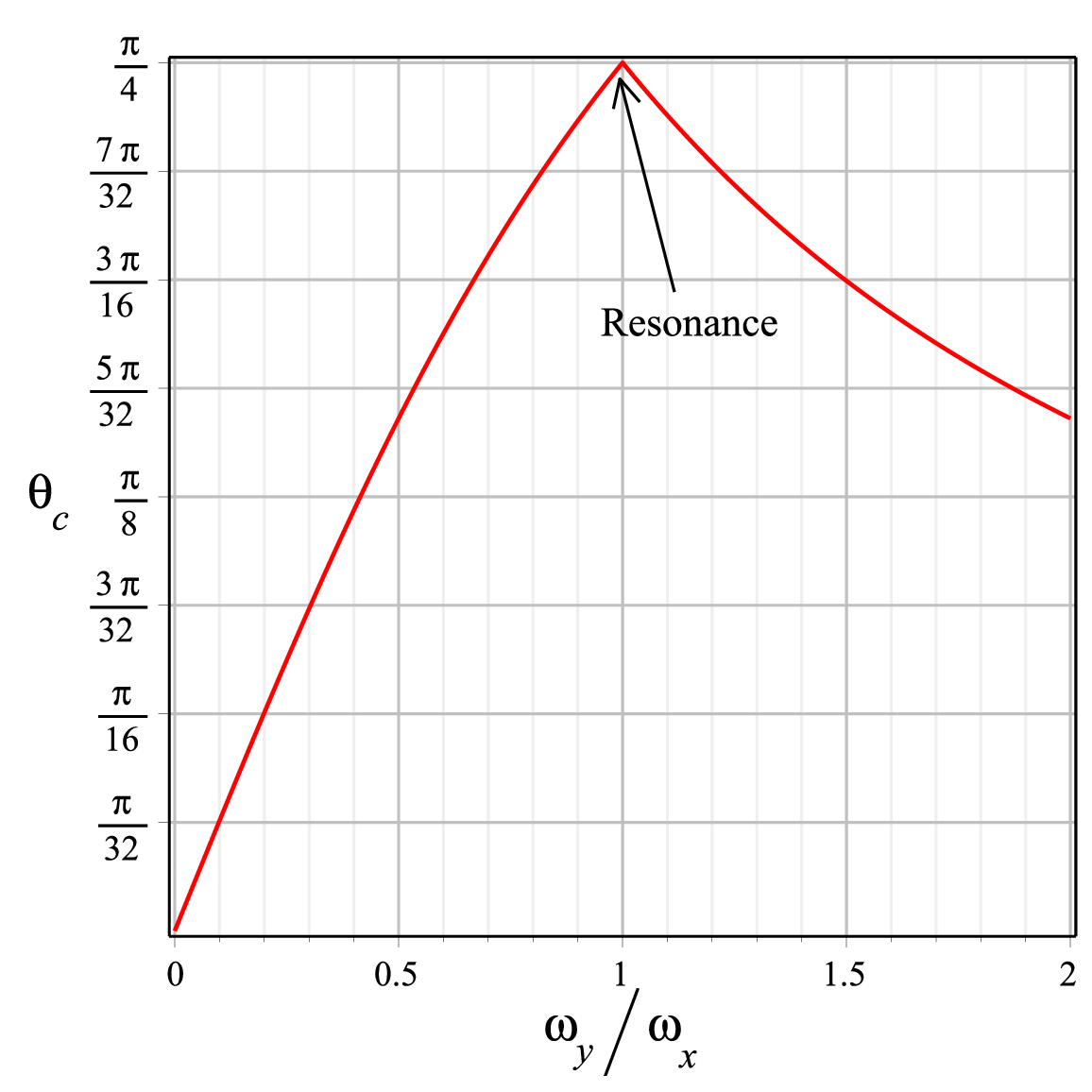}
	\caption{(color online) The evolution of $\theta_c$ versus resonance rate $r=\omega_y/\omega_x$. \label{fig1}}
\end{figure}

We analyze a physical system consisting of two coupled harmonic oscillators, where the interaction between them is represented by a coupling term of the form $\hat{x}\hat{y}$ \cite{ham1,ham2}. In particular, the Hamiltonian of the system is given by
\begin{eqnarray}
\hat{\mathbb{H}}&=&\frac{1}{2} \left(\hat{p}^2+\hat{q}^2\right)+\frac{1}{2}\omega_{x}^2\hat{x}^2+\frac{1}{2}\omega_{y}^2\hat{y}^2-\epsilon \hat{x}\hat{y} \label{Eq1}
\end{eqnarray}
where we consider hereafter $\hbar=m=1$, without loss of generality \cite{macedo}. The position and momentum operators satisfy the commutation relations $[\hat{x},\hat{p}]=[\hat{y},\hat{q}]=i$ and $[\hat{x},\hat{y}]=[\hat{q},\hat{p}]=0$. 
By rotating the coordinates as
\begin{align}
&\label{3}	\hat{X}= \hat{x}\cos\theta +\hat{y}\sin\theta , \quad \hat{Y}= -\hat{x}\sin\theta +\hat{y}\sin\theta \\
&	\hat{P}= \hat{p}\cos\theta +\hat{q}\sin\theta, \quad \hat{Q}= -\hat{p}\sin\theta+\hat{q}\sin\theta \label{3-3}
\end{align}
we transform the Hamiltonian \eqref{Eq1} into 
\begin{equation}\label{eq3}
\hat{\mathbb{H}}_d=\frac{1}{2} \left(\hat{P}^2+\hat{Q}^2\right)+\frac{1}{2}\vartheta^{2}_{x}\hat{X}^2
+\frac{1}{2}\vartheta^{2}_{y}\hat{Y}^2
\end{equation}
where the rotation angle $\theta$ is 
\begin{eqnarray}
\theta=\frac{1}{2}\arctan\left(\frac{2\epsilon}{\omega_x^2-\omega_{y}^2}\right)
\end{eqnarray}
and the normal frequencies are
\begin{eqnarray}
\vartheta_{x,y}^2=\frac{\omega_x^2+\omega_y^2}{2}\pm\frac{1}{2}\sqrt{(\omega_x^2-\omega_y^2)^2+4\epsilon^2}=\omega_{x,y}^2\pm\epsilon  \tan\theta \label{Eq5}
\end{eqnarray}
The eigenenergies of  (\ref{eq3}) are expressed as follows
\begin{eqnarray}
E_{(n,m)}=\frac{\vartheta_x}{2}(2n+1)+\frac{\vartheta_y}{2}(2m+1)
\end{eqnarray}
To ensure that the eigenvalues are real, we require that $\vartheta_{y}^2$ remains positive. This condition is satisfied by imposing the constraint $\epsilon < \omega_x \omega_y$. As a result, we  obtain a cut-off mixing angle $\theta_{c}$, such that 
\begin{eqnarray}
\theta_{c}= \frac{\text{sgn}(1-r)}{2}\arctan\left(\frac{2r}{1-r^2}\right)
\end{eqnarray}
 where $r=\frac{\omega_y}{\omega_x}$ and $\text{sgn}(x)=+1$ if $x\geq0$ and $0$ otherwise. When both oscillators are near resonance
$(r\to 1)$, the mixing angle 
$\theta_c=\displaystyle{\lim_{\epsilon \to \omega_x \omega_y}} \theta$ approaches $\frac{\pi}{4}$.  For different values of $r$ the mixing angle $\theta_c$ varies as shown in {\sf{Figure}}~\ref{fig1}. 
\noindent It is easy to obtain the associated eigenfunctions  as
\begin{align}
	\Psi_{(n,m)}(X,Y)&= \Psi_n(X)\otimes \Psi_{m}(Y)\\\notag
	&=\frac{1}{\sqrt{2^{n+m}n!m!}}\left(\frac{\vartheta_x\vartheta_y}{\pi^2}\right)^{\frac{1}{4}}e^{-\frac{\vartheta_x}{2}X^2}e^{-\frac{\vartheta_y}{2}Y^2}H_n(\sqrt{\vartheta_x}X)H_m(\sqrt{\vartheta_y}Y)
\end{align} 
where $H_n(x)$ are for Hermite polynomials, and the new variables ($X,Y$) are defined in (\ref{3}).

\section{Wigner function and phase space fluctuations\label{sec3}}

\subsection{Wigner fuction}

The Wigner function is a phase-space description of quantum states, providing a unique way to visualize and analyze quantum systems in both position and momentum space \cite{Wigner1}.  Its non-positivity makes it a crucial tool for understanding and characterizing quantum mechanics \cite{Wigner2, Wigner3}. We show that the Wigner function corresponding to the diagonalized Hamiltonian is separable 
\begin{eqnarray}\label{10}
W_{(n,m)}(X,P;Y,Q)= W_{n}(X,P)\times W_m(Y,Q)
\end{eqnarray}
where the involved functions are given by
\begin{align}
W_n(X,P)&=\frac{1}{\pi} \int d\mathbb{X}\ \Psi_{n}^{\ast}(X+\mathbb{X})\ \Psi_{n}(X-\mathbb{X})\ e^{2iP\mathbb{X}}\\\notag
&=\frac{(-1)^n}{\pi}e^{-\frac{1}{\vartheta_x}\left({\vartheta_x^{2}}X^2+P^{2}\right)}\mathcal{L}_{n}\left[\frac{2}{\vartheta_x}\left(\vartheta_x^{2}X^2+P^{2}\right)\right]
\\
W_m(Y,Q)&=\frac{1}{\pi} \int d\mathbb{Y}\ \Psi_{m}^{\ast}(Y+\mathbb{Y})\ \Psi_{m}(Y-\mathbb{Y})\ e^{2iQ\mathbb{Y}}\\\notag
&=\frac{(-1)^m}{\pi}e^{-\frac{1}{\vartheta_y}\left({\vartheta_y^{2}}Y^2+P^{2}\right)}\mathcal{L}_{m}\left[\frac{2}{\vartheta_y}\left(\vartheta_y^{2}y^2+P^{2}\right)\right]
\end{align}
and $\mathcal{L}_{n}(x)$ are for Laguerre polynomials. Using the variable changes (\ref{3}), we map the Wigner function \eqref{10} in terms of the original coordinates
\begin{align}
&W_{(n,m)}(x,p;y,q)=\frac{1}{\pi^2} (-1)^{n+m} e^{-{\vartheta_{x}}(x\cos\theta+y\sin\theta)^2-\frac{1}{\vartheta_x}(p\cos\theta+q\sin\theta)^2}\ e^{-{\vartheta_{y}}(x\sin\theta-y\cos\theta)^2-\frac{1}{\vartheta_y}(p\sin\theta-q\cos\theta)^2}\\
& \times\mathcal{L}_n\left({2\vartheta_{x}}(x\cos\theta+y\sin\theta)^2+\frac{2}{\vartheta_x}(p\cos\theta+q\sin\theta)^2\right) \times\mathcal{L}_m\left(2\vartheta_{y}(x\sin\theta-y\cos\theta)^2+\frac{2}{\vartheta_y}(p\sin\theta-q\cos\theta)^2\right)\label{wigner} \notag
\end{align}
This demonstrates the coupling between the two harmonic oscillators, a factor that will play a key role in the subsequent analysis.

\subsection{Phase space fluctuations}
Quantum fluctuations have profound implications in various quantum phenomena, including quantum tunneling, entanglement, and the evolution of quantum states in different potential landscapes. Characterizing and understanding these fluctuations is crucial for interpreting and predicting the behavior of quantum systems on microscopic scales. To go further, we recall that the expectation value of an operator $\mathcal{O}$ is given by  \cite{phase}
\begin{eqnarray}
\langle\mathcal{O}\rangle=\int_{\mathbb{R}^4}dx\ dy\ dp\ dq\
\mathcal{O} \ W_{(n,m)}(x,y,p,q) \label{average}.
\end{eqnarray}
As a result, we show the following expectation values for the positions and moments
\begin{align}
&\langle x^2\rangle =\frac{(1+2n)}{2\vartheta_x (1+\mu^2)}+\frac{(1+2m)\mu^2}{2\vartheta_y (1+\mu^2)}\\
& \langle y^2\rangle=\frac{(1+2n)\mu^2}{2\vartheta_x (1+\mu^2)}+\frac{(1+2m)}{2\vartheta_y (1+\mu^2)}
\\
&
\langle p^2\rangle =\frac{(1+2n)\vartheta_x}{2 (1+\mu^2)}+\frac{(1+2m)\vartheta_y\mu^2}{2(1+\mu^2)}\\
& \langle q^2\rangle= \frac{(1+2n)\mu^2\vartheta_x}{2 (1+\mu^2)}+\frac{(1+2m)\vartheta_y}{2(1+\mu^2)}
\\
&
\langle xy\rangle=\frac{\mu}{2(1+\mu^2)}\left(\frac{1+2n}{\vartheta_x}-\frac{1+2m}{\vartheta_y}\right)\\
&\langle pq\rangle=\frac{\mu}{2(1+\mu^2)}\left({(1+2n)}{\vartheta_x}-{(1+2m)}{\vartheta_y}\right)
\\
&\langle x^2 y^2\rangle= \frac{3\left[(1+2m(1+m)\vartheta_x^2+(1+2n(1+n)\vartheta_y^2\right]\mu^2+(1+2m)(1+2n)\vartheta_x\vartheta_y(\mu^4-4\mu^2+1)}{4\vartheta_x^2\vartheta_y^2(1+\mu^2)^2}
\\
&
\langle p^2 q^2\rangle=\frac{3\left[(1+2m(1+m)\vartheta_y^2+(1+2n(1+n)\vartheta_x^2\right]\mu^2+(1+2m)(1+2n)\vartheta_x\vartheta_y(\mu^4-4\mu^2+1)}{4(1+\mu^2)^2}
\\
&
\langle x^2q^2\rangle=\frac{1}{2}\left[\frac{(1+m+m^2+n+n^2)\mu^2}{(1+\mu^2)^2}\right]+\frac{(1+2m)(1+2n)(\mu^4\vartheta_x^2+\vartheta_y^2)}{4(1+\mu^2)^2 \vartheta_x\vartheta_y}
\\
&
\langle y^2p^2\rangle=\frac{1}{2}\left[\frac{(1+m+m^2+n+n^2)\mu^2}{(1+\mu^2)^2}\right]+\frac{(1+2m)(1+2n)(\mu^4\vartheta_y^2+\vartheta_x^2)}{4(1+\mu^2)^2 \vartheta_x\vartheta_y}
\end{align}
where $\mu=\tan\theta$ and $\vartheta_{x,y}$ are defined in (\ref{Eq5}). These results will be used in the next section to gain further insight into the present system.

\section{Heisneberg uncertainties and virtual excitations \label{sec4}}

Heisenberg uncertainty principle and quantum correlations intersect to deepen our understanding of quantum mechanics \cite{H}. The uncertainty principle states that we cannot simultaneously know the exact position and momentum of a particle, revealing a fundamental limit to measurement accuracy \cite{H1}. Quantum correlations, especially in entangled particles, show that measuring the state of one particle immediately affects its entangled partner, regardless of distance. In recent years, there has been a significant focus on the interplay between Heisenberg uncertainties and quantum correlations \cite{DP1,DP2,DP3,intro13}. In this context, for our system the phase space areas corresponding to the two oscillators are given by
\begin{align}
&[\mathcal{A}_x(n,m)]^2=\left[\Delta x\Delta p (n,m)\right]^2=\frac{1}{4}\frac{\left(\vartheta_x\mu^2(2m+1)+(2n+1)\vartheta_y\right)\left(\vartheta_y\mu^2(2m+1)+(2n+1)\vartheta_x\right)}{(\mu^2 +1)^2\vartheta_x\vartheta_y}
\\
&[\mathcal{A}_y(n,m)]^2=
\left[\Delta y\Delta q (n,m)\right]^2=\frac{1}{4}\frac{\left(\vartheta_x(2m+1)+\mu^2(2n+1)\vartheta_y\right)\left(\vartheta_y(2m+1)+\mu^2(2n+1)\vartheta_x\right)}{(\mu^2 +1)^2\vartheta_x\vartheta_y}.
\end{align}
We will now proceed with a detailed discussion. Specifically, if 
$\mu=1$, the oscillators are in resonance  and the both areas become equal 
\begin{eqnarray}
\mathcal{A}_x(n,m)=\mathcal{A}_y(n,m)\label{resphase}
\end{eqnarray}
and when the oscillators are decoupled ($\mu=0$), the uncertainty relations take the form 
\begin{eqnarray}
\mathcal{A}_x(n,m)=\frac{2n+1}{2}, \quad \mathcal{A}_y(n,m)=\frac{2m+1}{2}.
\end{eqnarray}
The uncertainties are independent of the physical parameters of the system. Moreover, the uncertainty associated with the oscillator in the $x$ (or $y$) direction depends only on its quantum excitation number $n$ (or $m$). Now, considering the case of weak couplings, $\epsilon \ll \omega_x, \omega_y$, the frequencies of the oscillators approach equality, i.e., $\vartheta_x \sim \vartheta_y$, leading to the following result
\begin{align}
&\mathcal{A}_x(n,m)=\frac{(2m +1)\mu^2+2n+1}{2(\mu^2+1)}\\
&\mathcal{A}_y(n,m)=\frac{(2n +1)\mu^2+2m+1}{2(\mu^2+1)}.
\end{align}
It is noteworthy that for the symmetric state, where $n=m$, the uncertainties are reduced to
\begin{eqnarray}
\mathcal{A}_x(n,m)= \mathcal{A}_y(n,m)=\frac{2n+1}{2}
\end{eqnarray}
which are independent of physical parameters.

\subsection{ Quantum virtual excitations}
To analyze the virtual excitation, we first introduce the expressions for the creation and annihilation operators associated with both oscillators. They are given by
\begin{align}
&(a_x^{\dagger})^{\dagger}=a_x=\sqrt{\frac{\omega_x}{2}}\hat{x}+\frac{i}{\sqrt{2\omega_x}}\hat{p}\\ &(a_y^{\dagger})^{\dagger}=a_y=\sqrt{\frac{\omega_y}{2}}\hat{y}+\frac{i}{\sqrt{2\omega_y}}\hat{q}
\end{align}
which satisfy the commutation relation $[a_i,a^\dagger_i]=\mathbb{I}$, while all other commutators vanish. These operators can be used to determine the expectation value of the excitation numbers $\langle a^{\dagger}_x a_x\rangle=\langle N_x\rangle$ and $\langle a^{\dagger}_y a_y\rangle=\langle N_y\rangle$ in the state $\Psi_{(n,m)}$. After straight algebras we get
\begin{align}
\langle  N_x\rangle &= \frac{1 }{4(\mu^2+1)}\left( \frac{\omega_x}{\vartheta_x}+\frac{\vartheta_x}{\omega_x}\right)(1+2n)+\frac{\mu^2 }{4(\mu^2+1)}\left( \frac{\omega_x}{\vartheta_y}+\frac{\vartheta_y}{\omega_x}\right)(1+2m)-\frac{1}{2}\\ 
\langle N_y \rangle &=\frac{\mu^2}{4(1+\mu^2)}\left( \frac{\omega_y}{\vartheta_x}+\frac{\vartheta_x}{\omega_y}\right)(1+2n)+\frac{1 }{4(\mu^2+1)}\left( \frac{\omega_y}{\vartheta_y}+\frac{\vartheta_y}{\omega_y}\right)(1+2m)-\frac{1}{2}.
\end{align}
At this stage we have some comments in order. In fact, for the weak coupling regime, the excitation numbers are reduced to the following
\begin{align}
&\langle  N_x\rangle=\frac{(2n+1)+\mu^2(2m+1)}{2(\mu^2+1)}-\frac{1}{2}\\
& \langle N_y\rangle=\frac{\mu^2(2n+1)+(2m+1)}{2(\mu^2+1)}-\frac{1}{2}.
\end{align}
These results show that the excitation numbers $\langle N_x \rangle$ and $\langle N_y \rangle$ are influenced by the quantum numbers $n$ and $m$ as well as by the parameter $\mu$ related to the coupling strength. Despite the weak coupling, there is a non-negligible interplay between the excitations of the two oscillators, reflecting the underlying correlation introduced by the coupling.
For the case $\mu = 1$ it is clearly seen that the excitations are equal to
\begin{align}
	\langle N_x \rangle &= \langle N_y \rangle = \frac{n + m}{2}.
\end{align}
Now, for $\mu = 0$, we observe the following behavior
\begin{align}
	\langle N_x \rangle = n, \quad
	\langle N_y \rangle = m.
\end{align}
Additionally, in the ground state ($n = m = 0$) the system has no excitations and the state becomes empty. This is expected, since the ground state corresponds to the absence of excitations in both oscillators, i.e,
\begin{equation}
	\langle N_x \rangle = \langle N_y \rangle = 0
\end{equation}
which reflects the fact that there is no energy stored in the system. However, if we consider more complex scenarios, such as ultra-strong coupling (USC), the situation changes because the condition
\begin{align}
	\frac{\omega_{x,y}}{\vartheta_{x,y}} + \frac{\vartheta_{x,y}}{\omega_{x,y}} > 2.
\end{align}
Even in the ground state, the oscillators can be populated with virtual excitations. This is due to the breakdown of the usual assumption that the ground state is empty, as discussed in \cite{intro14, virtual}. In the USC regime, the excitations are non-zero, leading to a modified expression for the excitation numbers
\begin{equation}
	\langle N_x \rangle, \langle N_y \rangle > 0.
\end{equation} 
This shows that the system can support virtual particles, even at the lowest energy level, which has important implications for quantum systems under strong interactions.

\section{Quantum entanglement \label{sec5}}

We start by recalling that Makarov in \cite{Makarov1} used the Schmidt decomposition to analyze the entanglement of two coupled oscillators. To explicitly determine the Schmidt modes $\lambda_k$, the assumption of weak coupling, $\epsilon\ll \omega_x, \omega_y$, is used, which leads to the approximation $\vartheta_x\sim\vartheta_y\sim\omega_x\sim \omega_y$. 
Consequently,  the $\lambda_k$ are obtained as follows
\begin{eqnarray}
\lambda_k(n,m)=\frac{\mu^{2(k+n)}m!n!}{(1+\mu^2)^{m+n}k!(m+n-k)!}\left(P_{n}^{(-(1+m+n),m-k)}\left(-\frac{2+\mu^2}{\mu^2}\right)\right)^2
\label{Makarov}
\end{eqnarray}
and therefore the corresponding purity is
\begin{equation}
	\mathbb{P}(n,m)=\sum\limits_{k=0}^{n+m}\lambda_{k}^2(n,m).
\end{equation}
It is easy to check that the purity of the ground state reduces to $\mathbb{P}(0,0)=1$, hence the state is separable. However, the marginal purity of the ground state for two coupled oscillators has already been shown in \cite{intro12, intro13,intro14}
\begin{eqnarray}
\mathbb{P}(0,0)=\left(1+\frac{\mu^2(\vartheta_x-\vartheta_y)^2}{{(1+\mu^2)^2}\vartheta_x\vartheta_y}\right)^{-\frac{1}{2}} .\label{ground}
\end{eqnarray}

\begin{figure}[H]
\centering
  \includegraphics[width=8cm, height=6cm]{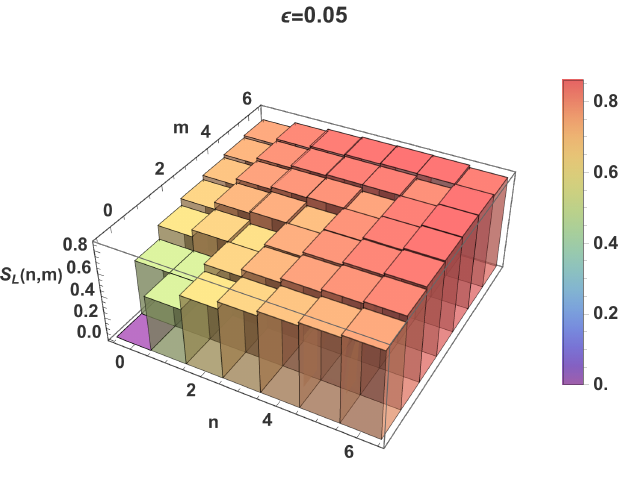}
\includegraphics[width=8cm, height=6cm]{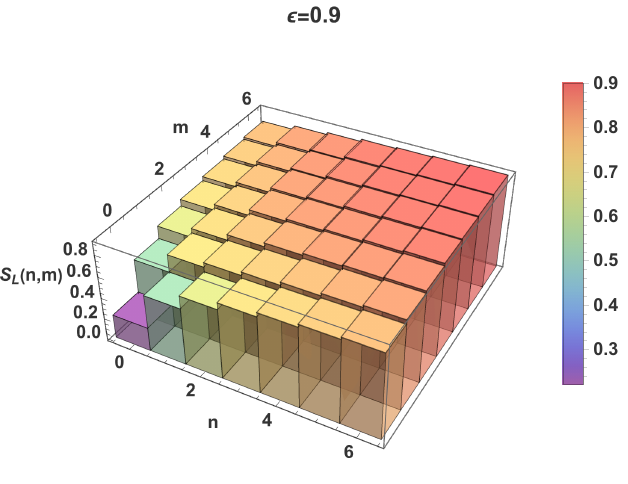}
\caption{(color online) Histograms show the evolution of the linear entropy $S_L(n,m)$ versus the quantum numbers $n$ and $m$ at resonance $\omega_x=\omega_y=1$, with two different values of the coupling $\epsilon$: $=0.05$ (weak coupling), $0.9$ (ultra-strong coupling) and . \label{fig22}}
\end{figure}

Inspired by recent advances in ultra-strong coupling physics \cite{Us1,Us2,Us3,Us4,Us5}, we aim to shift our focus to the computation of quantum entanglement in regimes beyond the weak coupling approximations adopted in \cite{Makarov1}. For this purpose, it is straightforward to check the purity of the global state, which can be expressed as 
\begin{equation}
4\pi^2\int_{\mathbb{R}^4}dx\ dy\ dp\ dq\ W^{2}_{(n,m)}(x,p,y,q)=1
\label{purity1}
\end{equation}
where $W_{(n,m)}(x,p,y,q)$ is given by (\ref{wigner}).
The purity provides insight into the entanglement between the subsystems, with a value of 1 indicating a pure state and values less than 1 indicating mixed states. This allows us to explore the behavior of quantum entanglement in the ultra-strong coupling regime.
Accordingly, entanglement can be assessed by calculating the marginal purities, specifically by evaluating one of the following purities
\begin{align}
	\mathbb{P}_x(n,m) &= 2\pi \int_{\mathbb{R}^2} dx \, dp \, W_{(n,m)}(x,p) \\
	\mathbb{P}_y(n,m) &= 2\pi \int_{\mathbb{R}^2} dy \, dq \, W_{(n,m)}(y,q)
\end{align}
where \( W_{(n,m)}(x,p) \) and \( W_{(n,m)}(y,q) \) are the marginal Wigner functions defined by
\begin{align}
	W_{(n,m)}(x,p) &= \int_{\mathbb{R}^2} dy \, dq \, W_{(n,m)}(x,p;y,q) \\
	W_{(n,m)}(y,q) &= \int_{\mathbb{R}^2} dx \, dp \, W_{(n,m)}(x,p;y,q)
\end{align}
which allow us to compute the purity in the phase space of each oscillator, which provides a means to quantify the degree of entanglement in the system. Furthermore, the global state is pure (see (\ref{purity1})), hence $\mathbb{P}_x(n,m)=\mathbb{P}_y(n,m)$.
Note that evaluating these integrals is not a straightforward task. To proceed, we use an approach based on the Rodrigues formula for Laguerre polynomials \cite{laguerre}
\begin{eqnarray}
	\mathcal{L}_n(x) &=& \frac{1}{n!} \frac{d^n}{du^n} \left( \frac{e^{-\frac{xu}{1-u}}}{1-u} \right) \Bigg\vert_{u=0}. \label{laguerre}
\end{eqnarray}
By integrating over \(y\) and \(q\) and using (\ref{laguerre}), we get the following purity expression
\begin{equation}
	\mathbb{P}_{(n,m)} = 
\frac{1}{(n! m!)^2} \frac{d^n}{dv^n} \frac{d^m}{dw^m} \frac{d^n}{du^n} \frac{d^m}{ds^m}	
	\left[ \frac{\prod\limits_{\kappa=u,s,v,w} \frac{2}{1-\kappa}}{\varpi(u,s,v,w,\vartheta_x,\vartheta_y) \varpi(u,s,v,w,{\vartheta_x^{-1}},{\vartheta_y^{-1}})} \right]_{u,s,v,w=0} \label{purity}
\end{equation}
with \(\varpi\) represented by
\begin{equation}
	\varpi(u,s,v,w,\vartheta_x,\vartheta_y) = \sqrt{f(u,s,\vartheta_x,\vartheta_y) \Omega(v,w,\vartheta_x,\vartheta_y) + f(v,w,\vartheta_x,\vartheta_y) \Omega(u,s,\vartheta_x,\vartheta_y)}
\end{equation}
and we have set up the following functions
\begin{align}
	f(u,s,\vartheta_x,\vartheta_y) &= \vartheta_x \vartheta_y \frac{u+1}{1-u} \frac{s+1}{1-s} \\
	\Omega(u,s,\vartheta_x,\vartheta_y) &= \frac{u+1}{1-u} \vartheta_x \frac{\mu^2}{1+\mu^2} + \frac{s+1}{1-s} \vartheta_y \frac{1}{1+\mu^2}.
\end{align}
To verify this method, we compute the purity for the ground state \(\Psi_{(0,0)}\), and it can be easily shown that \(\mathbb{P}(0,0)\) simplifies to the result found in (\ref{ground}).
To give a numerical illustration of quantum entanglement for different quantum numbers \(n\) and \(m\), we show in {\sf Figure} \ref{fig22} histograms of the linear entropy \(S_L(n,m)\) as a function of the quantum numbers \((n,m)\) at resonance \(\omega_x = \omega_y\) and for different coupling strengths \(\epsilon\). For simplicity, we normalize both frequencies to unity, so the coupling strength is \(\epsilon\in [0,1[\). For weak coupling \(\epsilon = 0.05\), the quantum entanglement increases with \(n\) and \(m\), while the ground state remains separable with \(S_L \sim 0\). As the coupling approaches the ultra-strong regime \(\epsilon = 0.9\), the ground states of both oscillators become entangled, and the excited states exhibit even higher levels of entanglement.
 
We now turn our attention to examining the differences between our exact results and those reported by Makarov in \cite{Makarov1}. It is important to note that Makarov's results are based on the Schmidt modes defined in  (\ref{Makarov}). By calculating the linear entropy \( S_L(n,m) \), which quantifies the entanglement for pure states, we can better understand the discrepancies between our exact results and those derived by Makarov. To quantify these differences, we introduce the trade-off quantity
\begin{equation}
	\Delta S_L(n,m) = S_L(n,m) - S_L^M(n,m)
	\label{deltaSL}
\end{equation}
where \( S_L^M(n,m) \) is the linear entropy associated with the Schmidt modes computed by Makarov, defined by
\begin{equation}
	S_L^M(n,m) = 1 - \sum_{k=0}^{n+m} \lambda_k^2(n,m)
	\label{SL_Makarov}
\end{equation}
and \( \lambda_k \) given by (\ref{Makarov}). Thus, \( S_L(n,m) \) corresponds exactly to our results.

\begin{figure}[H]
\centering
\includegraphics[width=8cm, height=6cm]{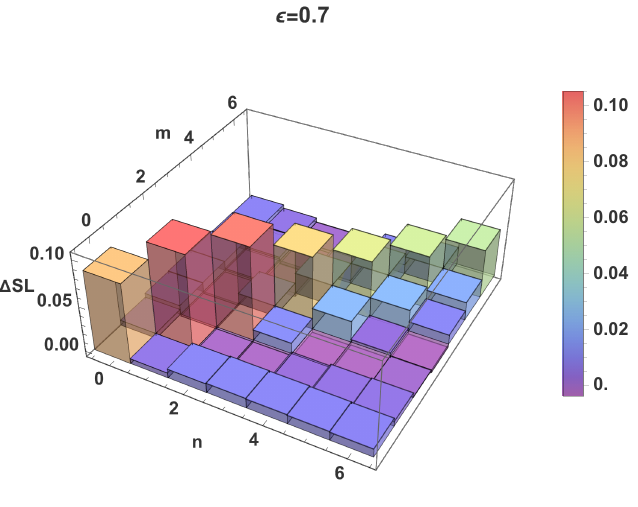}
\includegraphics[width=8cm, height=6cm]{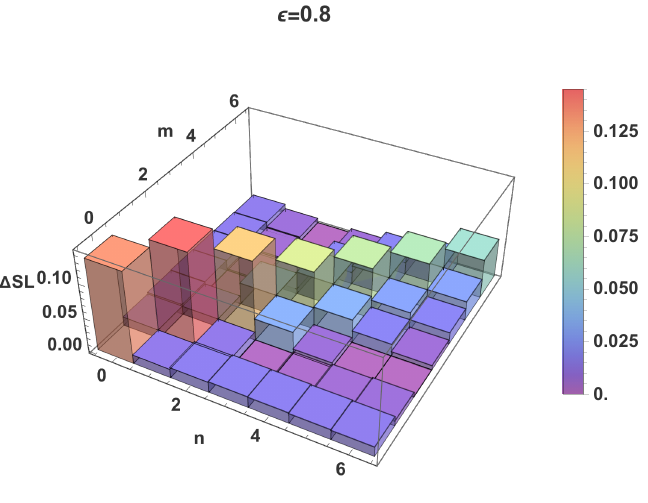}
\includegraphics[width=8cm, height=6cm]{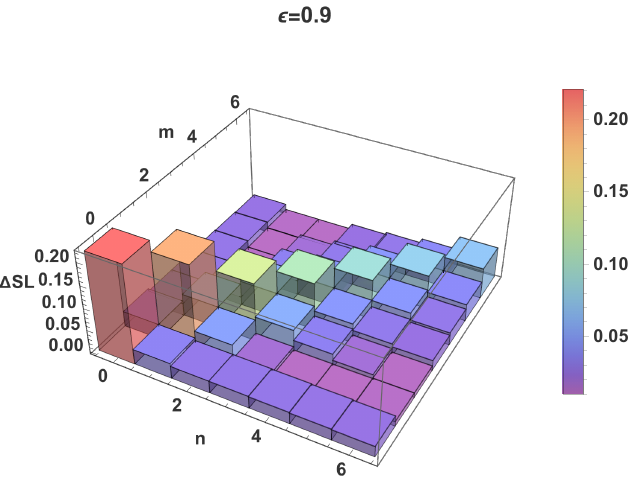}
\includegraphics[width=8cm, height=6cm]{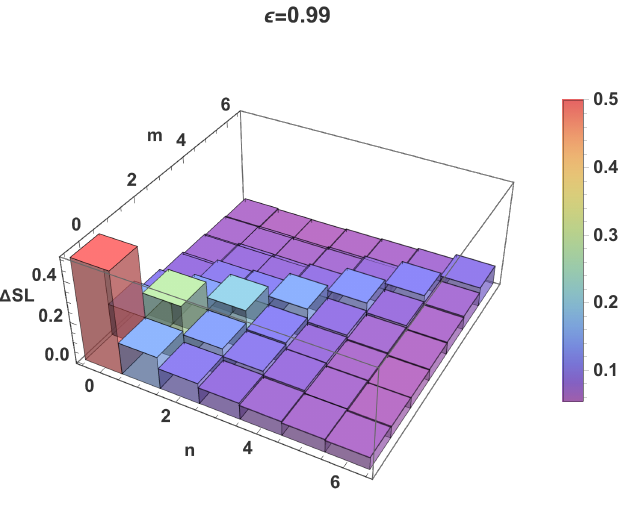}
\caption{(color online) Histograms illustrate the divergence of the entanglement, computed using Makarov's $S_L^M(n,m)$ and our exact results $S_L(n,m)$, as a function of the quantum numbers $n$ and $m$. These are shown for different values of the ultra-strong coupling $\epsilon \in \{0.7, 0.8, 0.9, 0.99\}$ and at resonance, where $\omega_x = \omega_y = 1$. \label{fig33}}
\end{figure}

In {\sf Figure} \ref{fig33}, we show the evolution of \( \Delta S_L(n,m) \) over different quantum numbers \( (n,m) \), with \( \omega_x = \omega_y = 1 \) and varying the coupling constant \( \epsilon \). It is important to note that Makarov's results, \( S_L^M(n,m) \) evaluated at resonance, show no dependence on \( \epsilon \), which is in stark contrast to our exact results (as shown in {\sf Figure} \ref{fig22}). In addition, we note that \( S_L^M(0,0) = 0 \) for all values of \( \epsilon \), a result that contradicts previous results, such as those in \cite{DP1, virtual}, where entanglement was found to be non-zero for \( (n,m) = (0,0) \). 
As expected, the discrepancies between Makarov's approximation and our exact results become particularly apparent for smaller quantum numbers and in the ultra-strong coupling regime. For example, \( \Delta S_L(n,m) \) tends to be more pronounced when \( n = m \), a pattern not captured by Makarov's model. This highlights a significant deviation in the behavior of entanglement, especially in regimes where the coupling constant \( \epsilon \) is large. In conclusion, while the Makarov approach provides some insight, it is clear that it is not universally applicable, as it deviates significantly from the exact results in several key aspects.

\section{Stationary  quantum steering \label{sec6}}
 \subsection{Steering and its quantification}

 	Quantum steering is a phenomenon in quantum mechanics in which one system, often called the "sender", can influence or "steer" another system, called the "receiver", in a way that transcends classical physics \cite{WisemanCavalcanti2011}. Unlike classical correlations or even entanglement, quantum steering allows for non-local influence that is instantaneous or faster than light, even when the systems are spatially separated \cite{AYHe20112011}. This ability to steer a system goes beyond what classical physics can explain and demonstrates a distinctive and stronger form of quantum correlation. Quantum steering plays a key role in understanding the non-local nature of quantum mechanics and has important implications for quantum communication and quantum information processing \cite{intro7}.

To detect and quantify quantum steering between two subsystems, denoted \( x \) and \( y \), we use a specific parameter derived from the inequality introduced in \cite{calvacanti}. This parameter provides a robust framework for assessing the degree of steerability in a quantum system. Quantum steering is a non-local effect where one subsystem (the "sender") can influence the state of another subsystem (the "receiver") in a way that cannot be explained by classical correlations. The steerability of the oscillator \( y \) from the oscillator \( x \) is quantified by the following expression
\begin{eqnarray}
	S_{x \to y}^{(n,m)} = \max\left[\left|\langle a_x a_y^\dagger \rangle\right|^2 - \left\langle a_y a_y^\dagger \left( a_x a_x^\dagger + \frac{1}{2} \right) \right\rangle, 0 \right]. \label{39}
\end{eqnarray}
Similarly, the steerability in the opposite direction, from \( y \) to \( x \), is computed by
\begin{eqnarray}
	S_{y \to x}^{(n,m)} = \max\left[\left|\langle a_x a_y^\dagger \rangle\right|^2 - \left\langle a_x a_x^\dagger \left( a_y a_y^\dagger + \frac{1}{2} \right) \right\rangle, 0 \right]. \label{40}
\end{eqnarray}
The key concept here is that the system is considered "steerable" in the direction \( x \to y \) (or vice versa) if and only if \( S_{x \to y}^{(n,m)} > 0 \) (or \( S_{y \to x}^{(n,m)} > 0 \)). This means that the state exhibits quantum steering if one subsystem can influence the other in a non-local way, as quantified by the parameters defined above.

It is worth making a comparison between quantum steering and entanglement. In fact, quantum steering differs from entanglement in important ways. While entanglement represents symmetric correlations between subsystems, quantum steering can exhibit asymmetry. Specifically, \( S_{x \to y}^{(n,m)} \neq S_{y \to x}^{(n,m)} \) indicates that the ability to steer is not necessarily mutual. This asymmetry means that one system (the sender) can control the state of the other (the receiver), but the reverse may not be true. 
In entanglement, the correlation between two subsystems is symmetric, i.e. the entanglement between system \( x \) and system \( y \) is the same as the entanglement between system \( y \) and system \( x \). However, quantum steering is inherently asymmetric. The non-local influence of one subsystem on another can occur in one direction, but not necessarily in the opposite direction. The sender can steer the state of the receiver, but this does not mean that the receiver can steer the state of the sender in return.

To quantify the asymmetry of quantum steering, we define the trade-off quantity \( \Delta S(n,m) \)
\begin{eqnarray}
	\Delta S(n,m) &=& \left|S_{x \to y}^{(n,m)} - S_{y \to x}^{(n,m)}\right|
\end{eqnarray}
This trade-off measure captures the degree to which the control is asymmetric. A value of \( \Delta S(n,m) = 0 \) would indicate that the control is symmetric, i.e., both subsystems can control each other equally. Conversely, a non-zero value of \( \Delta S(n,m) \) indicates that there is an imbalance between the controllability of the two subsystems. The larger the value of \( \Delta S(n,m) \), the more pronounced the asymmetry in the system.
Physically, asymmetry in quantum steering reflects a deeper property of quantum correlations. It implies that in certain quantum systems, one subsystem can have more influence or control over another subsystem.

\subsection{Weak coupling regime}
	
	In the context of weakly coupled harmonic oscillators, the coupling $\epsilon$ between the oscillators is assumed to be small relative to their frequencies $\omega_x$ and $\omega_y$. In this regime, the normal frequencies of the coupled system are approximately equal, i.e., $\vartheta_x \sim \vartheta_y \sim \omega_x \sim \omega_y \sim \omega$ \cite{Makarov1}. This approximation allows us to consider the system as essentially having a single characteristic frequency $\omega$, which simplifies the analysis of quantum correlations and steerability between the two subsystems.
	In this case, the quantum steering between the two oscillators in both directions, $(x \to y)$ and $(y \to x)$, reduces to a well-defined expression. Specifically, the steering quantifiers in both directions are related and have the same form, leading to the expression	
	\begin{eqnarray}
		S_{x \to y}^{(n,m)} = S_{y \to x}^{(m,n)} = \max\left[ -\frac{m + 2mn - (m+n)\mu^2 + (1+2m)n\mu^4}{2(1 + \mu^2)^2}, 0 \right]
	\end{eqnarray}	
	where $\mu = \tan \theta$ is a parameter characterizing the system, such as the angle between the quadratures of the oscillators. The form of this expression shows that the quantum steering between the two oscillators depends on both the quantum states of the oscillators and the specific coupling parameters.
	
	The steerability of the state $\Psi_{(n,m)}$—which is a generalized state describing the coupled system—depends on the values of the quantum numbers $n$ and $m$. In particular, we find that the system exhibits steerability in the $x \to y$ direction if and only if $n \neq 0$ and $m = 0$. Similarly, the system exhibits steerability in the $y \to x$ direction if and only if $m \neq 0$ and $n = 0$. This suggests a clear asymmetry in the steerability of the system based on the excitation levels of the two oscillators.	
	For a more explicit description of the quantum steering quantifiers, we find that for the case where $n \neq 0$ and $m = 0$, the quantum steering in the $x \to y$ direction (and similarly in the $y \to x$ direction for $m \neq 0$ and $n = 0$) reduces to the following simplified expression:	
	\begin{eqnarray}
		S^{(n,0)}_{x \to y} = S^{(0,n)}_{y \to x} = \frac{n \mu^2 (1 - \mu^2)}{2(1 + \mu^2)^2}.
	\end{eqnarray}
	This expression reveals several important features of the quantum steering behavior of the system. First, the quantum steering is an increasing function of the quantum number $n$ for a given value of $\mu$, meaning that higher excitation levels of the system lead to stronger steering effects. Moreover, since $n$ can take on arbitrarily large values, the quantum steering can in principle grow indefinitely, reaching infinite values for sufficiently high excitations. This is a remarkable feature of quantum systems, where the quantum correlations, in this case quantified by the steering, can diverge under certain conditions, a behavior also observed in quantum entanglement..
	This feature of infinite quantum steering is reminiscent of the phenomenon of entanglement in quantum mechanics, where the entanglement measures can also grow indefinitely as the system is pushed into higher excited states. However, an essential difference is that unlike entanglement, which can be symmetric in both directions (for a maximally entangled state), quantum steering exhibits a clear asymmetry: the ground state cannot steer an excited state, and steering is possible only in one direction.
	This result underscores the subtle differences between entanglement and steering, and highlights the role of quantum steering as a stronger form of quantum correlation that does not necessarily require entanglement. In addition, we conclude that quantum steering in weakly coupled harmonic oscillators is a highly non-symmetric phenomenon: it is only possible under certain conditions (i.e., with a non-zero excitation level in one oscillator and a zero excitation level in the other).
	Thus, the quantum steering behavior in this system provides a novel insight into the nature of quantum correlations and opens the door to further studies of steerability in more complex, interacting quantum systems.

\begin{figure}[H]
	\centering
		\includegraphics[width=8cm, height=6cm]{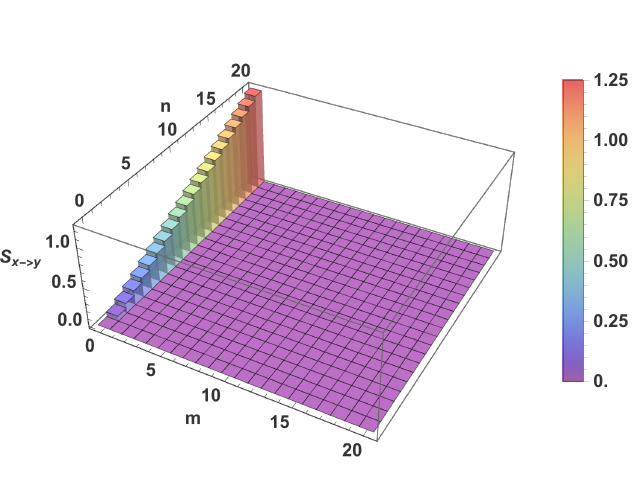}
\includegraphics[width=8cm, height=6cm]{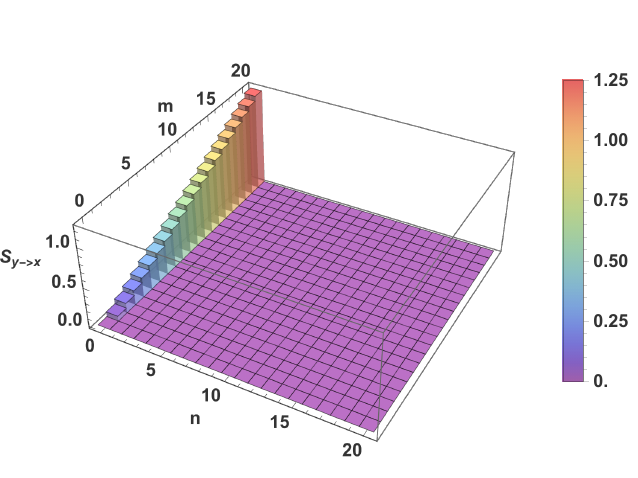} 
\caption{(color online)  The quantum steerings $S_{x\to y}^{(n,m)}$ and $S_{y\to x}^{(n,m)}$ versus the quantum numbers $n$ and $m$ for $\omega_x=\omega_y=1$ and $\mu=\frac{\sqrt{3}}{3}$.}\label{figure 2}
\end{figure}

For a clear visual representation of the steering dynamics, we present in {\sf Figure}~\ref{figure 2} the histograms of the steering quantifiers $S_{x \to y}^{(n,m)} = S_{y \to x}^{(n,m)}$ for different quantum numbers $(n,m)$ and with the coupling parameter $\mu = \frac{\sqrt{3}}{3}$.  In particular, the histograms show that the steering between successive quantum states is separated by a fixed value of $\frac{1}{16}$, which serves as the maximum difference in steering between adjacent states. This separation illustrates how the steering quantifiers change discretely as a function of the quantum numbers $n$ and $m$.
It is important to note the complete asymmetry in the steering behavior, where the product of the steering quantities in both directions, $S_{x \to y}^{(n,m)} \times S_{y \to x}^{(n,m)}$, always vanishes. This asymmetry underscores the unique nature of quantum steering in this system, where steering can only occur in one direction: either from $x \to y$ or from $y \to x$, but not both simultaneously for the same quantum state.
Furthermore, the figure shows two critical points where quantum steering vanishes completely. First, when $\mu = 1$, which corresponds to the resonance condition, the steering between the oscillators disappears completely. This is because at resonance, the interaction between the oscillators leads to maximum symmetry, which effectively cancels out the steering effect. Second, when the oscillators are decoupled ($\mu = 0$), no steering is possible because the absence of coupling prevents any quantum influence between the subsystems. These two limit cases serve as boundaries for the steering behavior of the system and are essential for understanding the full range of dynamics in weakly coupled harmonic oscillators.

\subsection{Ultra-strongly coupled regime}

The ultra-strong coupling regime refers to a scenario in quantum optics and condensed matter physics where the coupling strength between light (photons) and matter (typically atoms, molecules, or other quantum systems) becomes so large that it is comparable to or exceeds the intrinsic transition frequencies of the system \cite{USCrev}. This regime transcends the more commonly studied weak and strong coupling regimes and gives rise to novel and exotic physical phenomena \cite{USCrev, Us1, Us2, Us3, Us4, Us5}. In our analysis of quantum steering in this ultra-strong coupling regime, we first assume that both oscillators have identical intrinsic frequencies, $\omega_x = \omega_y = 1$, which can be done without loss of generality. Under this assumption, the normal modes of the coupled system are characterized by the normal frequencies
\begin{equation}
	\vartheta_{x,y} = \sqrt{1 \pm \epsilon}
	\label{eq:norm_freq}
\end{equation}
where $\epsilon$ is the coupling strength. In addition, the mixing parameter between the oscillators becomes $\mu = 1$. With this setup it is easy to show that the quantum steering vanishes in both directions
\begin{equation}
	S_{x\to y}^{(n,m)} = S_{y\to x}^{(n,m)} = 0, \quad \forall (n,m,\epsilon).
	\label{eq:steering_zero}
\end{equation}
This result implies that in the resonant case ($\mu = 1$), steerability between the oscillators is completely unattainable, regardless of the excitation levels or coupling strengths. The reason for this lies in the symmetry of the system: at resonance, the phase space areas of the oscillators are equal. In particular, the areas occupied by the quantum states in phase space, $\mathcal{A}_x(n,m)$ and $\mathcal{A}_y(n,m)$, are identical
\begin{equation}
	\mathcal{A}_x(n,m) = \mathcal{A}_y(n,m)
	\label{resphase}
\end{equation}
which eliminates the possibility of quantum steering between the two subsystems. This holds even when one oscillator is in the ground state and the other is highly excited, since the resonance condition enforces this strict symmetry. We conclude that when two oscillators are in resonance, their phase space areas become equal, leading to a complete suppression of quantum steering in both directions.


To explore the conditions under which quantum steering can be restored, we consider scenarios in which the system is detuned from resonance. When the resonance condition is broken, quantum steering can emerge depending on the excitation levels of the oscillators. In particular, when the system is slightly detuned from resonance, the steering quantifiers no longer vanish. We find the following conditions for non-zero steering
\begin{align}
	S_{x\to y}^{(n,m)} &\neq 0, \quad \forall n \neq 0,  m = 0 \label{eq:steering_x}\\
	S_{y\to x}^{(n,m)} &\neq 0, \quad \forall m \neq 0,  n = 0. \label{eq:steering_y}
\end{align}
These results highlight the asymmetry of quantum steering in detuned systems: steering is only possible when one of the oscillators is excited while the other remains in the ground state. The absence of steering in the resonant case ($\mu = 1$) can be understood as a direct consequence of the equal phase space areas, which vanish as soon as the symmetry between the oscillators is broken by detuning.

\begin{figure}[H]	
 \centering
\includegraphics[width=7cm, height=5cm]{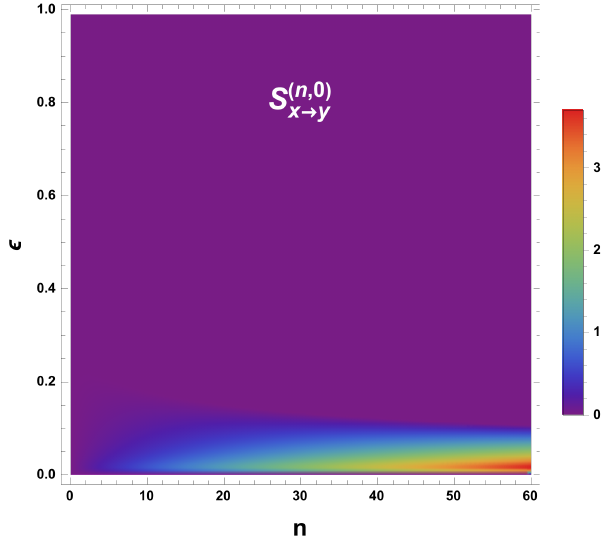}
\includegraphics[width=7cm, height=5cm]{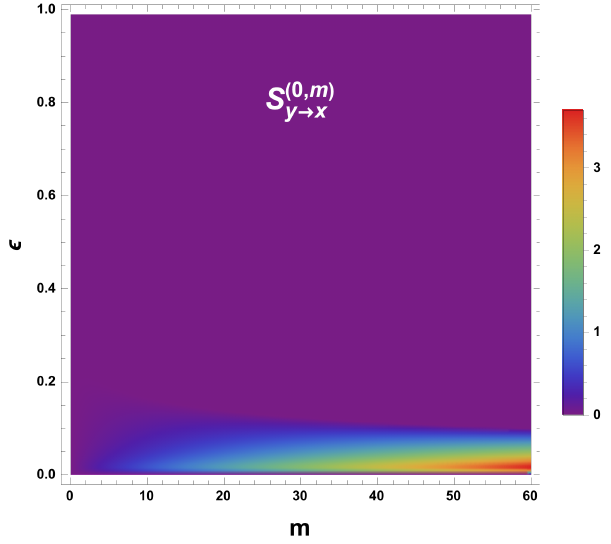} 
\caption{(color online)   The quantum steering in the two directions $S_{x\to y}^{(n,0)}$ and $S_{y\to x}^{(0,m)}$ versus the coupling $\epsilon \in [0,\omega_y]$ for $\omega_x=1$  and $\omega_y=0.99$.}\label{fig3}
\end{figure}

{\sf Figure} \ref{fig3} shows the effect of a small frequency detuning from resonance by setting $\omega_x = 1$ and $\omega_y = 0.99$. This small detuning allows us to observe how the quantum steering is revived in both directions, $S_{x\to y}^{(n,m)}$ and $S_{y\to x}^{(n,m)}$, as a function of the quantum numbers $(n,m)$ and the coupling strength $\epsilon$. The plotted results show that even a small deviation from resonance is sufficient to restore the steering between the oscillators. As the quantum number $(n,m)$ increases, the quantum steering becomes more pronounced, indicating that higher energy states contribute significantly to the steerability. This suggests that the excitation plays a critical role in determining the extent of non-local quantum correlations in the system. In addition, we observe that ultra-strong coupling suppresses steering between weakly excited states. In particular, as $\epsilon$ approaches the ultra-strong coupling regime, steering between low quantum number states becomes increasingly suppressed. This suppression highlights the delicate interplay between coupling strength and energy levels in determining quantum steerability. In summary, this analysis demonstrates two key results: (1) small frequency detuning can revive quantum steering that is otherwise nullified at resonance, and (2) while higher excited states exhibit enhanced steering, ultra-strong coupling can significantly reduce steerability between low excited states. These results highlight the complex dynamics of quantum steering under different coupling and detuning conditions.

The steerability is also highly sensitive to the coupling strength between the oscillators. In the regime of weak coupling, the quantum steerability is essentially zero, indicating no observable steering effect. As the coupling strength is increased, the steerability between the oscillators increases, showing a clear correlation between coupling and quantum steering. However, this relationship is not linear. When the coupling reaches very high values, the steerability begins to decrease and eventually disappears. This non-monotonic behavior implies that there is an optimal range of coupling strengths where quantum steering is maximized. It is important to note that this effect is not observed for entanglement, which instead increases monotonically with the coupling parameter.

\begin{figure}[H]
	\centering
	\includegraphics[width=7cm, height=5cm]{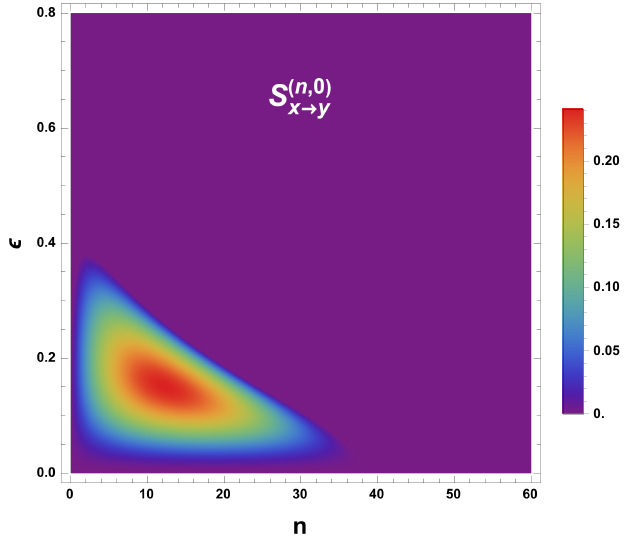}
	\includegraphics[width=7cm, height=5cm]{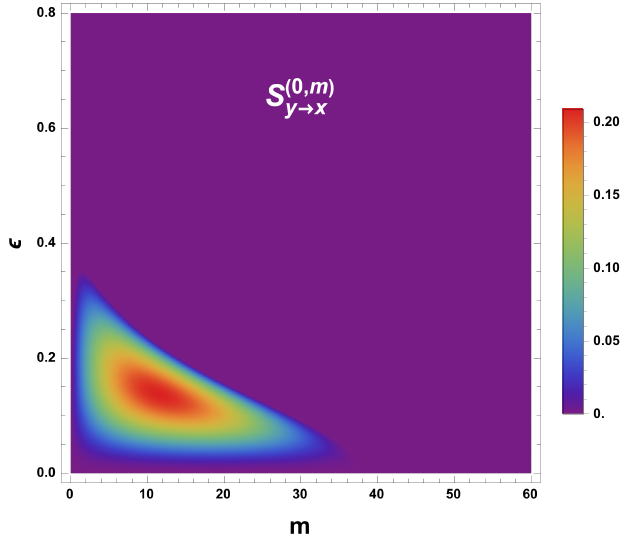}
	\caption{(color online)   The quantum steering in the two directions $S_{x\to y}^{(n,0)}$ and $S_{y\to x}^{(0,m)}$ versus the coupling $\epsilon \in [0,\omega_y]$ for $\omega_x=1$  and $\omega_y=0.8$.}\label{fig4}
\end{figure}

In {\sf Figure} \ref{fig4}, we fix all parameters except the angular frequency of the $y$-oscillator and set $\omega_y = 0.8$. We observe a significant reduction in steerability in both directions, with the maximum steerability $\max\left[S_{x\to y}^{(n,m)}; S_{y\to x}^{(n,m)}\right] \sim 0.2$. In addition, we note that for highly excited states, the steering is completely lost, even in the case of high ultra-strong coupling, $\epsilon \to \sqrt{\omega_x \omega_y}$. This suggests that beyond a certain coupling threshold, quantum steering is strongly suppressed and is no longer observable for states with large quantum numbers.

\begin{figure}[H]
\centering
\includegraphics[width=7cm, height=5cm]{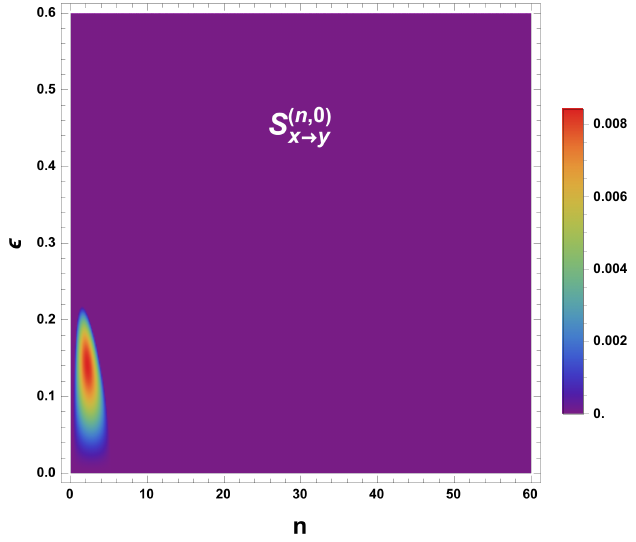}
\includegraphics[width=7cm, height=5cm]{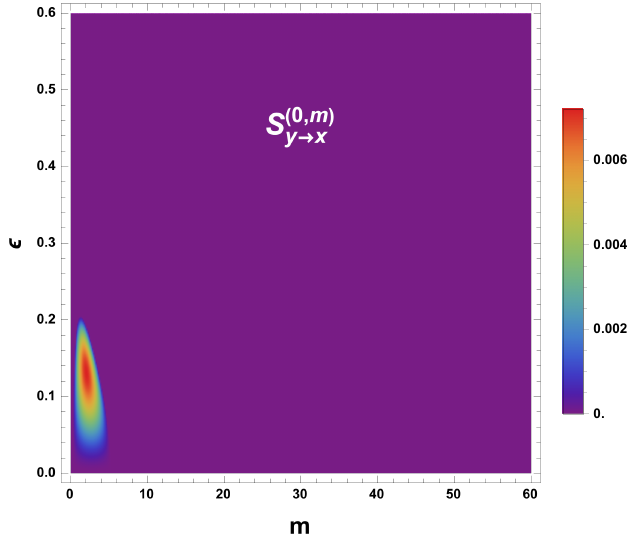}
\caption{(color online)   The quantum steering in the two directions $S_{x\to y}^{(n,0)}$ and $S_{y\to x}^{(0,m)}$ versus the coupling $\epsilon \in [0,\omega_y]$ for  $\omega_x=1$  and $\omega_y=0.6$.\label{fig5}}
\end{figure}

In {\sf Figure} \ref{fig5}, when the frequency $\omega_y$ is further reduced to $0.6$ while keeping the other parameters constant, we observe a strong suppression of the steerability in both directions, $x \to y$ and $y \to x$, even for low-excited states. This significant frequency mismatch disrupts the energy transfer and quantum correlations between the oscillators, leading to a rapid loss of quantum steering. Such a drastic reduction in steerability indicates that resonance or near-resonance conditions are crucial for maintaining quantum correlations, and large deviations from resonance greatly reduce the system ability  to exhibit steering.

\section{Conclusion\label{sec7}}

We thoroughly investigated quantum steering and quantum entanglement between two coupled harmonic oscillators, using the Wigner function in phase space instead of the Schmidt decomposition. We derived expressions for the purity $\mathbb{P}{(n,m)}$ and quantum steering in both directions, $S_{x \to y}^{(n,m)}$ and $S_{y \to x}^{(n,m)}$, leading to several significant selection rules.
Motivated by the study of the ultra-strongly coupled regime in two coupled harmonic oscillators and the seminal work of Makarov \cite{Makarov1}, we have computed the entanglement properties for general coupling scenarios. Our analysis shows that Makarov's results, originally presented for weak coupling regimes, do not hold universally. Specifically, we show that his results diverge significantly from our exact calculations. In particular, Makarov's results predict a separable ground state, which is in striking contrast to established results in the field. Our results thus highlight the limitations of Makarov's approach and provide a more accurate description of quantum entanglement in the ultra-strong coupling regime.

Consequently, we have studied and analyzed quantum steering within the system for general coupling parameters. We found that quantum steering is completely nullified between two excited oscillators, even when they are in the ultra-strong coupling regime. This indicates that despite the ultra-strong interaction, the effect of quantum steering does not emerge for excited states within the system. Similarly, our results show that quantum steering is absent for resonant oscillators. Moreover, we observed that the ground states themselves cannot steer any quantum state, indicating that they are not capable of inducing steering effects. In addition, we found that quantum steering becomes more pronounced near the resonance $(r \to 1)$ and within certain intervals of ultra-strong couplings. As a result, we showed that quantum steering is maximally asymmetric, i.e. there is no steering in at least one direction.

Our results provide a deeper understanding of how quantum steering and entanglement behave in coupled harmonic oscillators, emphasizing the role of excitation levels and coupling strength in influencing these quantum phenomena. The present results highlight quantum oscillators as promising candidates for quantum communication and advanced quantum technologies. Our findings reveal specific conditions under which quantum steering and entanglement can be optimized, providing valuable insights for improving quantum communication protocols and designing more effective quantum technologies. The ability to control and manipulate these quantum properties could lead to significant advances in quantum state transfer, error correction, and the development of precision quantum sensors and computing systems.

\end{document}